# Gate-Tunable Tunneling Resistance in Graphene/Topological Insulator Vertical Junctions

Liang Zhang[1,†], Yuan Yan[1,†], Han-Chun Wu[2], Dapeng Yu[1,3,4] & Zhi-Min Liao[1,3,*]

[1]State Key Laboratory for Mesoscopic Physics, School of Physics, Peking University, Beijing 100871, P.R. China

[2]School of Physics, Beijing Institute of Technology, Beijing, 100081, P.R. China

[3]Collaborative Innovation Center of Quantum Matter, Beijing, China

[4]Electron Microscopy Laboratory, School of Physics, Peking University, Beijing 100871, P. R. China

[†] These authors contributed equally to this work.

* Address correspondence to liaozm@pku.edu.cn

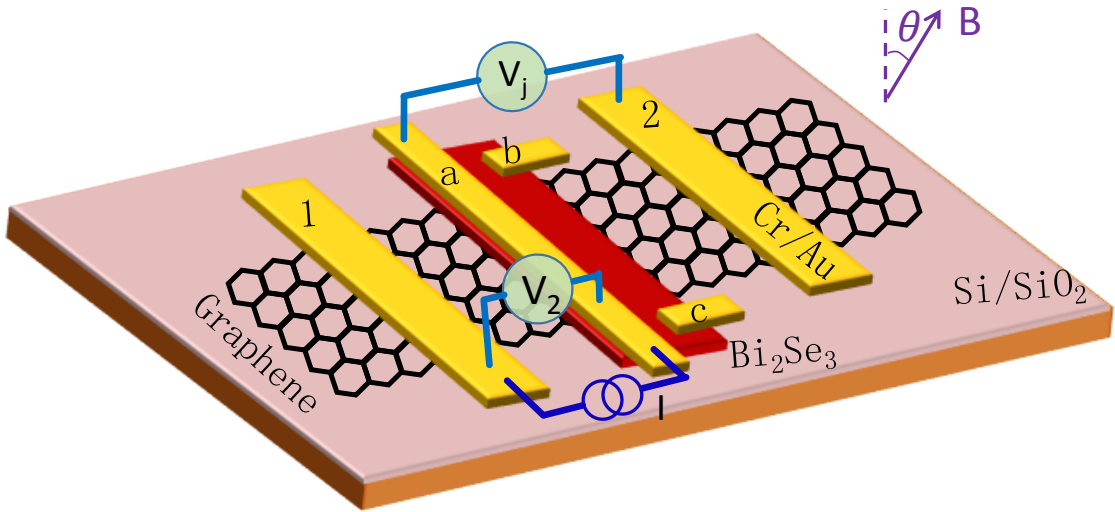

**ABSTRACT Graphene-based vertical heterostructures, particularly stacks incorporated with other layered materials, are promising for nanoelectronics. The stacking of two model Dirac materials, graphene and topological insulator, can considerably enlarge the family of van der Waals heterostructures. Despite well understanding of the two individual materials, the electron transport**

**properties of a combined vertical heterojunction are still unknown. Here we show the experimental realization of a vertical heterojunction between $Bi_2Se_3$ nanoplate and monolayer graphene. At low temperatures, the electron transport through the vertical heterojunction is dominated by the tunneling process, which can be effectively tuned by gate voltage to alter the density of states near the Fermi surface. In the presence of a magnetic field, quantum oscillations are observed due to the quantized Landau levels in both graphene and the two-dimensional surface states of $Bi_2Se_3$. Furthermore, we observe an exotic gate-tunable tunneling resistance under high magnetic field, which displays resistance maxima when the underlying graphene becomes a quantum Hall insulator.**

As model Dirac materials, graphene[1,2] and topological insulators,[3,4] have attracted vast attention due to their unique physical properties. Graphene based van der Waals heterojunctions, formed by stacking graphene with other materials, have been a prosperous avenue of research, which exhibit many interesting physical phenomena, including Coulomb drag of massless fermions,[5] metal-insulator transitions[6] and enhancement of spin-orbit coupling.[7] The transport of Dirac fermions through potential barriers in graphene has been revealed to demonstrate the well-known Klein tunneling.[8-10] Meanwhile, grapheme-based vertical devices have shown promising properties for Schottky diode and tunneling junction applications.[11-19] Through tuning the tunneling barrier, a high on-off conductance ratio in graphene field-effect tunneling transistors has been achieved.[11] Another type of Dirac material, bismuth selenide ($Bi_2Se_3$), known as a prototype topological insulator, possesses conducting surface states (SS) protected by the time-reversal symmetry.[3-4] Recently, theories have predicted significant modification of energy band structures induced by a proximity effect in graphene-topological insulator hybrid systems.[20-23] However, although the graphene-$Bi_2Se_3$ heterojunctions have been fabricated by vapor-phase deposition[24] and molecular beam epitaxy,[25,26] the vertical electronic transport properties in such heterostructures remain unstudied.

For $Bi_2Se_3$ grown on graphene, it is difficult to fabricate multi-terminal electrodes separately on graphene and $Bi_2Se_3$ to form the vertical transport devices. Fortunately, layer-by-layer stacking of graphene has been demonstrated to be an effective method to construct graphene-based vertical devices.[27] Here we report on the fabrication of the graphene-$Bi_2Se_3$ vertical devices using layer-by-layer stacking of graphene and $Bi_2Se_3$ nanoplates and the gate-tunable electronic transport in such vertical junctions. A cross-bar measurement configuration allows characterization of both the in-plane transport properties of the individual building blocks and the vertical transport properties across the graphene/$Bi_2Se_3$ interface. Exotic gate-modulated tunneling spectra are observed as the Fermi level of graphene is tuned to across the Landau levels (LLs) under high magnetic field. There are maxima of tunneling resistance as graphene is being a quantum Hall insulator (QHI) state. Due to the close relationship between the tunneling resistance and the position of Fermi level, the graphene-$Bi_2Se_3$ vertical junctions are promising for the attainment of fascinating electronic states in these Dirac materials.

**RESULTS AND DISCUSSION**

**Device Configurations.** The hybrid devices were fabricated by transferring high quality $Bi_2Se_3$ nanoplates onto monolayer graphene flakes, followed by the deposition of patterned Cr/Au (50/170 nm) electrodes (**Figure S1**). As shown in **Figure 1a**, the applied bias current $I$ flows from the bottom graphene (Electrode 1) to the upper $Bi_2Se_3$ (Electrode a). Three different voltages, $V_2$ between electrodes 1-a, $V_j$ between electrodes a-2, and $V_{xy}$ between electrodes b-c, are measured simultaneously. The resistance $R_j = V_j/I$ represents the junction resistance between graphene and $Bi_2Se_3$, as $V_j$ is measured using a cross bar-like configuration. The resistance $R_2 = V_2/I$ consists of the junction resistance and the graphene resistance in series. Therefore, the resistance $R_g = R_2 - R_j$ is the in-plane resistance of the graphene. Considering $V_{xy}$ is measured from the paired electrodes on $Bi_2Se_3$, $R_{xy} = V_{xy}/I$ represents the Hall-like resistance of $Bi_2Se_3$. A back gate voltage $V_g$ was applied to the $SiO_2$/Si (highly doped) substrate to tune the carrier density. The angle $\theta$ between the direction of the magnetic field and the $c$ axis of $Bi_2Se_3$ can be tilted by rotating the sample holder.

The side view schematic of the device demonstrates clearly the arrangement of the basic components and the equivalent circuit (**Figure 1b**), consisting of the excess graphene and the graphene-$Bi_2Se_3$ junction. **Figure 1c** illustrates the band diagram of the vertical graphene/$Bi_2Se_3$ junction without $V_g$ and bias voltage. The work function of intrinsic graphene is about 4.6 eV,[18] while the work function of the $Bi_2Se_3$ nanoplates after exposure to atmospheric conditions is approximately 4.0 eV.[28] A potential barrier forms at the interface of the junction with an unintentional tunneling layer, arising from the outside oxide layer on the $Bi_2Se_3$ nanoplates (~2 nm in thickness) and tiny space between graphene and $Bi_2Se_3$ mainly produced by the ripples in graphene.[29,30] The right upper panel in **Figure 1c** indicates the Dirac bands of $Bi_2Se_3$ and graphene are centered at the Γ point and K, K′ points in Brillouin zone, respectively.

**Structure Characterizations.** The fabricated devices were systematically characterized by transmission electron microscopy (TEM), energy dispersive x-ray spectroscopy (EDS), atomic force microscopy (AFM), and Raman spectrum. **Figure 2a** presents a high resolution TEM (HRTEM) image of a $Bi_2Se_3$ nanoplate. The lattice spacing is ~0.21 nm, which agrees with the interplanar spacing of [11$\bar{2}$0] planes. The EDS spectrum collected on a $Bi_2Se_3$ nanoplate in the TEM indicates the Bi to Se atomic ratio is ~2 : 3 (**Figure 2b**). The Cu signals in the EDS spectrum come from the TEM copper grid for holding samples. The optical image of a typical device is shown in **Figure 2c**. The white dashed lines mark the edges of the bottom graphene. **Figure 2d** shows the AFM height profile of a typical graphene-$Bi_2Se_3$ device along the red line indicated in the inset of the AFM topographic image. The thicknesses of the $Bi_2Se_3$ nanoplate and the top electrode are ~135 nm and ~218 nm, respectively. The Raman spectrum in **Figure 2e** excited by a 514 nm laser clearly indicates the nature of monolayer graphene, as the 2D peak to G peak intensity ratio is larger than 3 : 1. The Raman spectrum of a $Bi_2Se_3$ nanoplate is shown in **Figure 2f**. The pronounced two peaks located at ~131 cm$^{-1}$ and ~174 cm$^{-1}$ are assigned to $E_g^2$ and $A_{1g}^2$ phonon vibrational modes, respectively, which is consistent with the former report.[31]

**Transport through the Graphene-$Bi_2Se_3$ Interface.** The Cr/Au electrodes form ohmic contacts with both $Bi_2Se_3$ and graphene; however there is a potential barrier at the graphene/$Bi_2Se_3$ interface (**Figure S2**). **Figure 3a** shows the junction resistance $R_j$ increases with decreasing temperature, while the graphene resistance $R_g$ shows a very weak temperature dependence (See supporting **Figure S3** for more R-T curves measured from other samples). Apparently, the junction resistance $R_j$ shows different features in the high temperature region and low temperature region. At high temperatures, the over-barrier thermionic current dominates. The number of thermally excited carrier can be described by $n(T) \approx e^{-\Delta/k_BT}$, where $\Delta$ is the barrier height, and $k_B$ is the Boltzmann's constant. In other words, $R_j$ is exponentially sensitive to the reciprocal of temperature.[12] Thus the Arrhenius plot of $ln(R_j)$ *vs* $1/T$ is presented at high temperature (**Figure 3b**), giving $\Delta \sim 0.35$ meV (*i.e.* ~4 K). This fitting result is quite consistent with the raw R-T curve, which shows a dramatic increase of $R_j$ below 4 K. At low temperatures (T < 4 K), the tunneling current is determined by the density of states (DOS) in the graphene ($D_G$) and in the $Bi_2Se_3$ ($D_{TI}$), which can be expressed as[11,12,17,32]

$$I(V) \propto \int D_G(E) \cdot D_{TI}(E - eV) \cdot T(E) \cdot [f(E - eV) - f(E)] \cdot dE,$$

where *T(E)* is the tunneling probability and the function $f(E)$ is the Fermi-Dirac distribution function. In the low temperature region, the integral energy can be restricted to the range between $\mu$ and $\mu$ + e$V$, where $\mu$ is the chemical potential. Because of the large momentum mismatch between graphene and $Bi_2Se_3$ Fermi surface (See upper right panel of **Figure 1c**), phonon assisted tunneling (*i.e.* inelastic tunneling) should occur at the interface. Thus the power law $R_j \propto T^{-\alpha}$ is employed to fit the experimental data in this tunneling region and we obtain $\alpha \sim 0.15$ (**Figure 3c**). The phonon provides momentum to inject an electron from $Bi_2Se_3$ to the $K, K'$ points of graphene. Because the energy of the optical phonon mode in $Bi_2Se_3$ is only ~7.7 meV,[33,34] the phonons with energy up to 200 meV in graphene could be responsible for the inelastic tunneling in the graphene-$Bi_2Se_3$ junctions.[35] For an example, the 68 meV phonon mode at $K, K'$ points in graphene has been concluded to assist the tunneling in graphene/BN heterostructures.[36]

**Quantum Oscillations of the Tunneling Resistance.** The tunneling resistance is highly related to the density of states in graphene ($D_G$), Bi$_2$Se$_3$ ($D_{TI}$), and the potential barrier ($T(E)$). **Figure 4a** shows the junction resistance $R_j$ as a function of the reciprocal of perpendicular component of the magnetic field, $1/B_\perp = 1/(B\cos\theta)$, at 1.5 K. Quantum oscillations of $R_j$ are observed and all the peaks line up for the different angles $\theta$, indicating the 2D nature of these oscillations. To clarify the origins of the quantum oscillations, the Hall-like resistance $R_{xy}$ of the Bi$_2$Se$_3$ as a function of $1/(B\cos\theta)$ was measured and is shown in **Figure 4b**. The notable oscillations in $R_{xy}$ demonstrate an equal spacing of $B\cos\theta$ between the neighboring peaks and valleys, which can be ascribed to Shubnikov-de Hass (SdH) oscillations in the Bi$_2$Se$_3$ nanoplates. Although the tunneling resistance is expected to be related to the vertical transport across the junction, the 2D natures of both graphene and Bi$_2$Se$_3$ make the quantized Landau levels sensitive to the perpendicular magnetic field. Therefore, the $R_j$ is highly dependent on $B_\perp$. For comparison, the oscillations of resistance $\Delta R_j$ and $\Delta R_{xy}$ were extracted by subtracting the background, as shown in **Figure 4c**. It is found that the peak and valley positions of $\Delta R_j$ are consistent with that of $\Delta R_{xy}$ under high negative magnetic field. While, the peak and valley positions of $\Delta R_j$ are inverted with that of $\Delta R_{xy}$ under high positive magnetic field. This can be ascribed to the anti-symmetry of $\Delta R_{xy}$ in the presence of a magnetic field measuring through the Hall-like configuration between the electrodes b and c on Bi$_2$Se$_3$ nanoplates (**Figure S4**). To further reveal the influence of graphene on the quantum oscillations of $\Delta R_j$, the transfer curve of the whole bottom graphene layer was measured using electrodes 1 and 2 (**Figure S5**), where the Dirac point is located at $V_g \sim 5\,V$. Accordingly, the carrier concentration of the graphene at zero gate voltage is estimated to be $n_G = 3.6 \times 10^{11}\,cm^{-2}$. All the carriers can be coalesced into the lowest Landau level (LLL) in graphene when the magnetic field is higher than 4 T according to $n_G = geB/h$, where $e$ is the elementary charge, $h$ is Planck's constant and $g = 4$ for graphene. The magnetoresistance (MR) of the graphene flake, measured between electrodes 1 and 2, shows a large positive MR without oscillating features under high magnetic field, as the Fermi level enters into the LLL (**Figure 4d**). Therefore, the quantum oscillations of $\Delta R_j$ under high magnetic field are from the quantized landau levels of the surface states on the bottom surface of the $Bi_2Se_3$ nanoplate, suggesting that such a tunneling junction can be used to detect the DOS characteristics of $Bi_2Se_3$. Under relatively small magnetic fields, the oscillations of $\Delta R_j$ are mixed with the contributions from both $Bi_2Se_3$ and graphene.

The $R_j$ vs $B$ curves measured at different temperatures are shown in **Figure 4e**. The quantized Landau levels induced oscillations of $R_j$ are clearly observable at temperatures below 30 K. The quantum oscillations disappear at temperatures higher than 50 K, as a result of the thermal broadening of the Landau levels. Additionally, the temperature dependence of the $R_{xy}$-$B$ behavior in **Figure 4f** is quite similar to that of $R_j$. The Landau level fan diagram extracted from $R_{xy}$ indicates a berry phase of $\beta = 0.8\pi$ (**Figure S6**), suggesting the Dirac fermions of surface states dominate the transport properties of $Bi_2Se_3$.

**Gate-Tunable Tunneling Resistance.** The Fermi level of graphene can be tuned by a back gate voltage $V_g$, which offers a route to modify the tunneling resistance of the graphene-$Bi_2Se_3$ junction. The resistance $R_2$ and $R_j$ as a function of $V_g$ at 2 K is shown in **Figure 5a**. The $V_g$ dependent resistance $R_g$, reflecting the field-effect properties of the bottom graphene, shows a clear Dirac point (**Figure 5a**), which is highly consistent with the transfer curve of the underlying graphene flake measured between electrodes 1 and 2 (**Figure S5**). The Hall-like $R_{xy}$ under low magnetic field at different $V_g$ shows the carrier density of $Bi_2Se_3$ ($n_{TI}$) is not sensitive to the back gate voltage (**Figure S7**) due to the screening effect from the underneath graphene. Furthermore, the clear asymmetry behavior of $R_j$ as sweeping $V_g$ from negative to positive (**Figure 5a**) demonstrates different tunneling barriers as the Fermi level in graphene is tuned from valance band to conduction band. The transfer curves of $R_j$ and $R_g$ under an in-plane magnetic field of 14 T (**Figure S8**) are very similar to that at B=0 (**Figure 5a**), indicating that the in-plane magnetic field hardly affects the transport properties, which is mainly due to the 2D natures of both graphene and the surface states of $Bi_2Se_3$.

One of the exotic phenomena that emerge from the graphene-$Bi_2Se_3$ junctions is the characteristic of $R_j$-$V_g$ relationship under a perpendicular magnetic field of 14 T, as shown in **Figure 5b**. Pronounced peaks of $R_j$ occur at particular gate voltages, such as -6 V and 17.5 V. Specifically, the $R_j$ peaks are exactly located at the gate voltage positions with the $R_g$ valleys, that is, the tunneling resistance reaches a maximum as the graphene resistance decreases to a minimum. To figure out the evolution of the tunneling resistance with both magnetic field and gate voltage, the MR curves of $R_j$ at 2 K and for different gate voltages are investigated (**Figure 5c**). The MR (defined as MR=[R(B)-R(0)]/R(0)) at 2 K, is about 400% under B=14 T and $V_g$ = 17.5 V, while there is no obvious increase in the MR with increasing magnetic field up to 14 T for $V_g = \pm 60\ V$. This gate tunable tunneling resistance $R_j$ is due to the change of Fermi level position in graphene and the almost pinned Fermi level in $Bi_2Se_3$. The available DOS increases greatly as the Fermi level in graphene is tuned far away from the Dirac point using $V_g$ (**Figure 5d**), resulting in a decrease in the tunneling resistance.

Under high magnetic fields, the formation of Landau levels in graphene is shown in **Figure 6a**. Under a perpendicular magnetic field, quantum Hall states emerge in graphene,[1,2] resulting in large available DOS at each empty LL for tunneling electrons. On the other hand, there are no available tunneling states when the chemical potential $\mu_G$ of graphene resides in the energy gaps of the LLs. In such a case, graphene acts like an insulator, known as a QHI. The two main $R_j$ peaks at 17.5 V and -6 V (**Figure 6b**) are due to the fully occupied LLL for the electrons and holes in graphene, respectively.

At a fixed gate voltage, the available DOS in graphene can also be increased by enhancing the degeneracy of the LLs with increasing magnetic field. As the Fermi level locates at a certain LL in graphene under high magnetic fields, further increase of magnetic field can also reduce $R_j$ (**Figure S9**). At higher temperatures (>50 K), thermal broadening of LLs and the contribution of a thermionic current further diminish $R_j$ (**Figure S10**). The semimetal-QHI transition in graphene largely modifies the tunneling resistance in the graphene-$Bi_2Se_3$ junctions, suggesting new possibilities for probing the DOS in 2D materials.

**DISCUSSION**

The elaborate characterizations of resistance $R_{xy}$ and $R_g$ can provide rich information about the tunneling behavior. For vertical junctions of stacked 2D materials, in-plane transport naturally exists.[27] The carrier density $n_{TI}$ and the magnetic field induced Landau levels in $Bi_2Se_3$ can be plainly revealed by measuring the Hall-like $R_{xy}$ (**Figure 4** and **Figure S7**). Under high magnetic fields, the tunneling process is dominated by the DOS of $Bi_2Se_3$ as the Fermi level of graphene is near the Dirac point and will be in the LLL. Therefore, the graphene-based tunneling devices can be used as electronic states detectors. Here the gate voltage $V_g$ mainly tunes the carrier concentration $n_G$ in the underlying graphene and shifts the chemical potential by $\mu_G = \pm \hbar v_F (\pi |n_G|)^{1/2}$, where Fermi velocity $v_F$ is $\sim 10^6$ $m/s$. The Fermi level in our pristine graphene is located at ~ 0.07 eV below the Dirac point. The modulation limit of $\Delta \mu_G$ is $\sim \pm 0.24$ eV in our $V_g$ range ($\pm 60$ V). In graphene, the unevenly spaced Landau level energy can be expressed by $E_n = sgn(n)\sqrt{2e\hbar v_F^2 |n| B}$ in the presence of a perpendicular magnetic field, where $n$ is the LL index with a positive sign for electrons and negative one for holes.[1,2] Thus, we can further check the proposed explanation of the $R_j$ peaks quantitatively. To fill the n $= \pm 1$ LLs after full occupation of the lowest Landau level (n = 0), $\Delta V_g$ ($|V_g^{n=\pm 1} - V_g^{n=0}|$) should be ~19 V, theoretically. Then another two smaller peaks in $R_j$ would be expected to appear at gate voltages about 19 V away from 17.5 V and -6 V, respectively. This prediction is quite consistent with the fact that two secondary peaks are located at about 39 V and -25.8 V, as shown in **Figure 5b**. Although the vertical transport properties derived from the 2D Dirac bands do show the signatures for Dirac fermions through analyzing the quantum oscillations, the graphene-$Bi_2Se_3$ junctions with two distinct Dirac flavors do warrant further investigations.

**CONCLUSIONS**

In conclusion, we have studied the tunneling behavior in graphene-$Bi_2Se_3$ vertical junctions. The utilization of $R_{xy}$ and $R_g$, deducing the electronic states of the $Bi_2Se_3$ nanoplate and graphene monolayer, offers a feasible route to investigate and understand the tunneling resistance. The graphene-$Bi_2Se_3$ junctions exhibit a 2D nature and Dirac features in the tunneling transport properties. Exotic tunneling processes occur as graphene turns to be a quantum Hall insulator with the application of magnetic field and gate voltage. Moreover, graphene-based junctions could also be employed as DOS detectors.

**MATERIALS AND METHODS**

**Device Fabrication.** The $Bi_2Se_3$ nanoplates and graphene were grown by chemical vapor deposition method. The fabrication processes of the graphene-$Bi_2Se_3$ heterostructures are illustrated in **Figure S1**. First, the as-grown graphene was coated by a thin PMMA layer and patterned as microstamp *via* e-beam lithography (EBL) and $O_2$ plasma etching. An intentional hole with ~2 μm in diameter on each microstamp was used to transfer the graphene sheet onto Si substrate with 300 nm thickness $SiO_2$ using a micro-manipulator guided by an optical microscope. Regular marks were pre-fabricated on the substrate. The PMMA on graphene was dissolved in acetone. Second, a $Bi_2Se_3$ nanoplate was transferred on the top of the graphene sheet by a micro-manipulator. Third, standard EBL techniques were used to pattern the electrodes. Cr/Au (50 nm/170 nm) electrodes were deposited by electron beam evaporation. After a lift-off process, the device was finally completed.

**Transport Measurements.** The transport measurements of the $Bi_2Se_3$/graphene devices were performed in an Oxford cryostat with a variable temperature insert and superconductor magnet. The temperature can be decreased to 1.4 K and the magnetic field can be swept up to 14 T. The electrical results were obtained *via* standard low-frequency lock-in techniques. The bias current was 0.1 μA unless otherwise stated.

*Conflict of Interest:* The authors declare no competing financial interests.

*Supporting Information Available:* The Supporting Information is available free of charge on the ACS Publications website at DOI: 10.1021/acsnano.****.

Fabrication processes of the devices, current-voltage curves, resistance-temperature curves, Hall resistance, transfer curves of graphene, Landau fan diagram, $R_{xy}$ under low magnetic field, gate voltage and magnetic field modulated tunneling resistance (PDF)

*Acknowledgement.* This work was supported by MOST (Nos. 2013CB934600, 2013CB932602) and NSFC (Nos. 11274014, 11234001, 11327902).

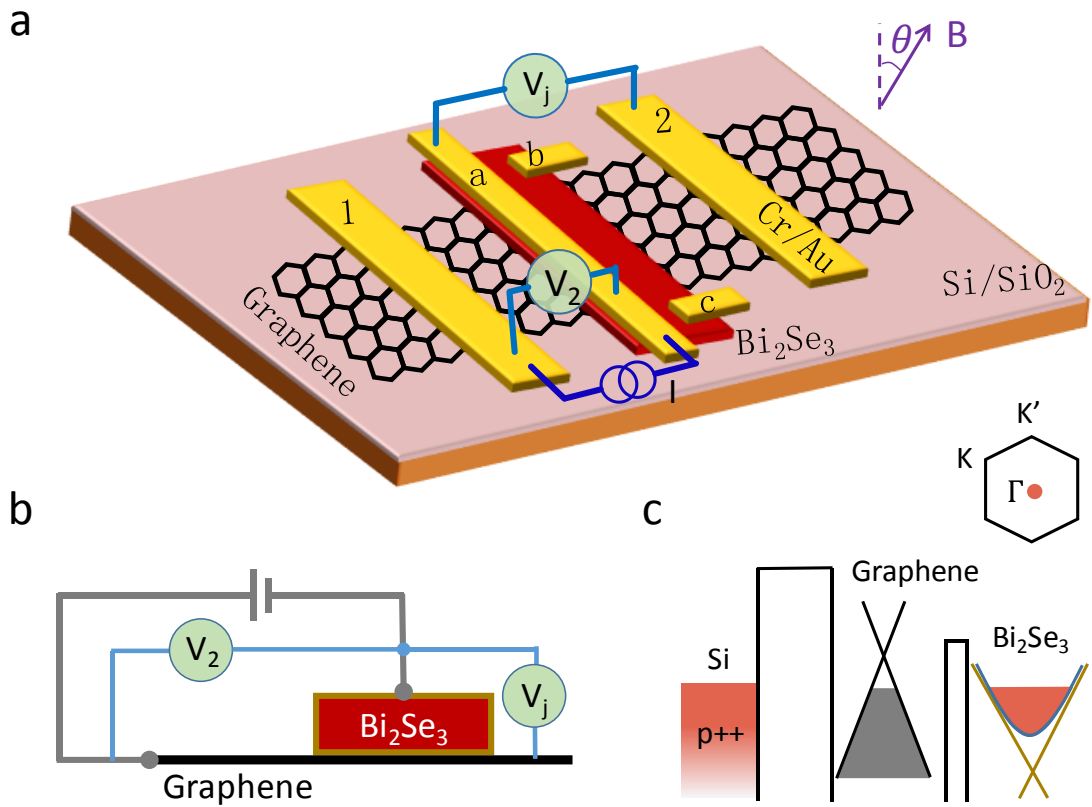

**Figure 1 | Device structure and diagrams.** (a) Schematic of graphene-$Bi_2Se_3$ heterostructure. Graphene is contacted with the electrodes 1 and 2. $Bi_2Se_3$ is contacted *via* the electrodes a, b, and c. The bias current is applied through electrodes 1 and a. (b) Side view of the measurement configuration. (c) Band diagram of the junction without bias voltage. The upper right panel shows the Brillouin zone of the graphene-$Bi_2Se_3$ hybrid system. The center of the Dirac bands of graphene (or $Bi_2Se_3$) is located at $K, K'$ (or $\Gamma$) points.

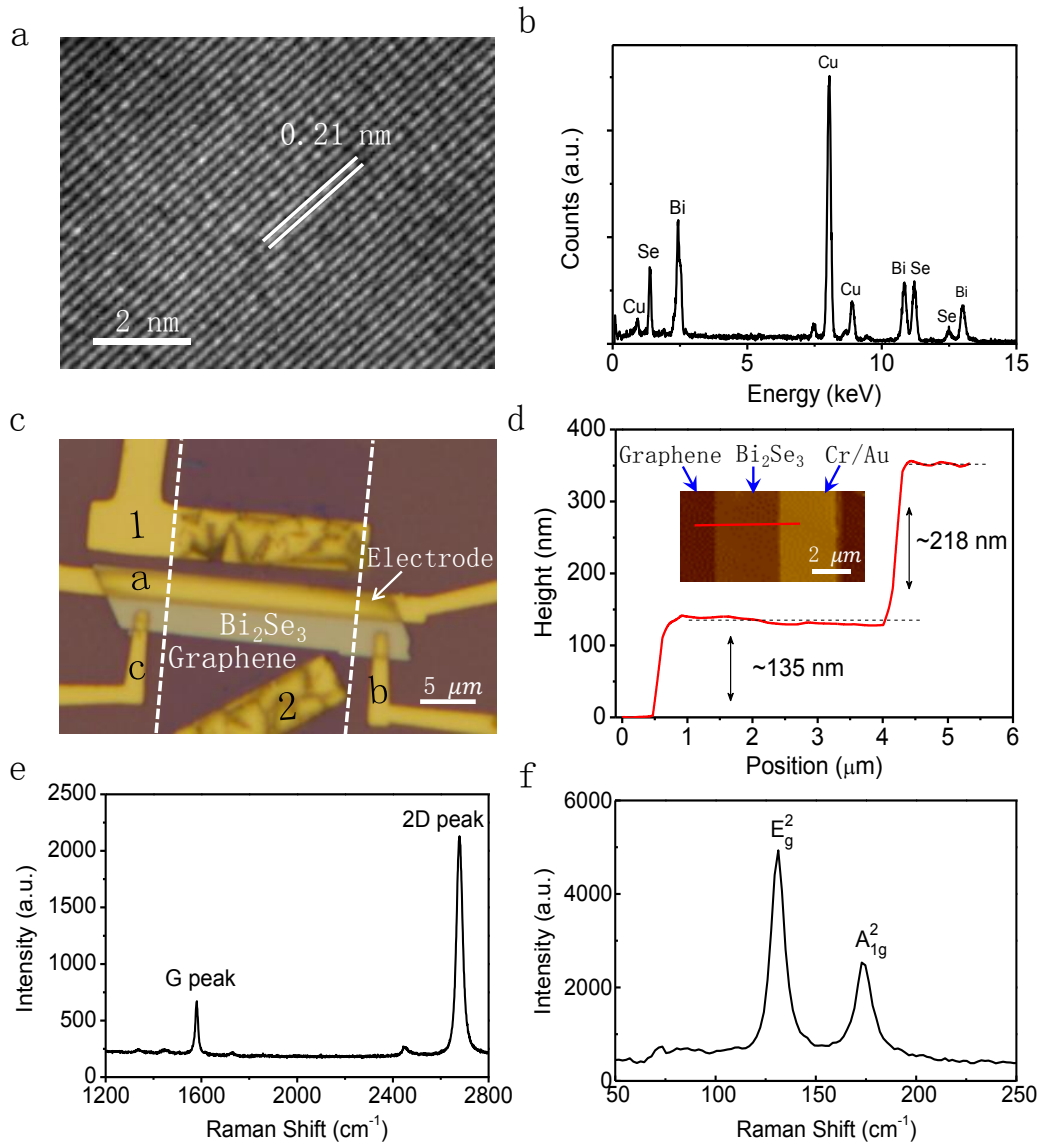

**Figure 2 | Characterizations of the devices.** (a) The HRTEM image of a $Bi_2Se_3$ nanoplate. The lattice spacing ~0.21 nm agrees with the interplanar spacing of $[11\bar{2}0]$ planes. (b) The EDS spectrum of a $Bi_2Se_3$ nanoplate. The Cu signals come from the TEM copper grid for holding samples. (c) Optical image of a typical multi-probe device. The white dashed lines mark the edges of the bottom graphene. Graphene is directly contacted with the electrodes 1 and 2. $Bi_2Se_3$ is contacted *via* the electrodes a, b, and c. (d) AFM height profile along the red line indicated in the inset of the AFM topography image. The thicknesses of the $Bi_2Se_3$ nanoplate and the top electrode are ~135 nm and ~218 nm, respectively. (e) Raman spectrum of the bottom graphene indicates the monolayer nature. (f) Raman spectrum of a $Bi_2Se_3$ nanoplate.

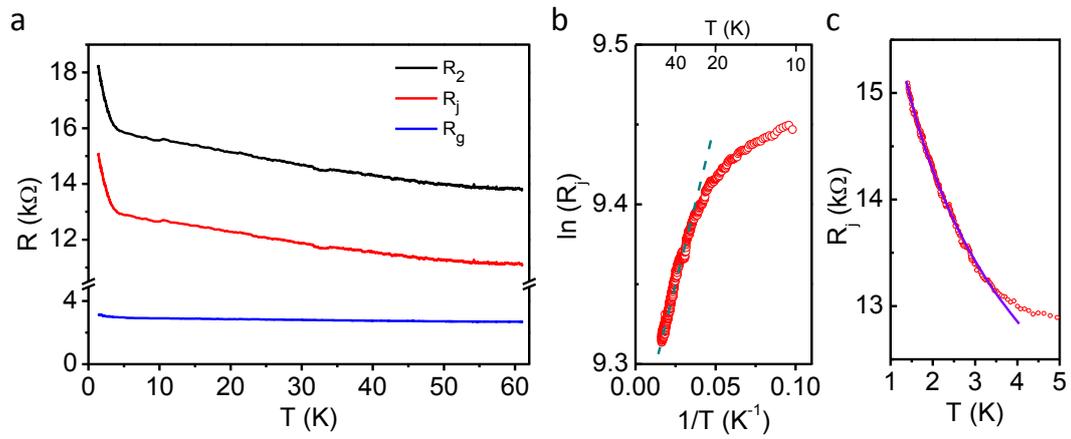

**Figure 3 | Temperature dependent resistance.** (a) $R_2$ (black line), $R_j$ (red line) and their difference $R_g$ (blue line) as a function of temperature. (b) Arrhenius plot of $R_j$ at high temperatures. (c) Power law fitting using $R_j \propto T^{-\alpha}$ at low temperatures, giving α~0.15. The dashed line in (b) and solid line in (c) are the fitting results.

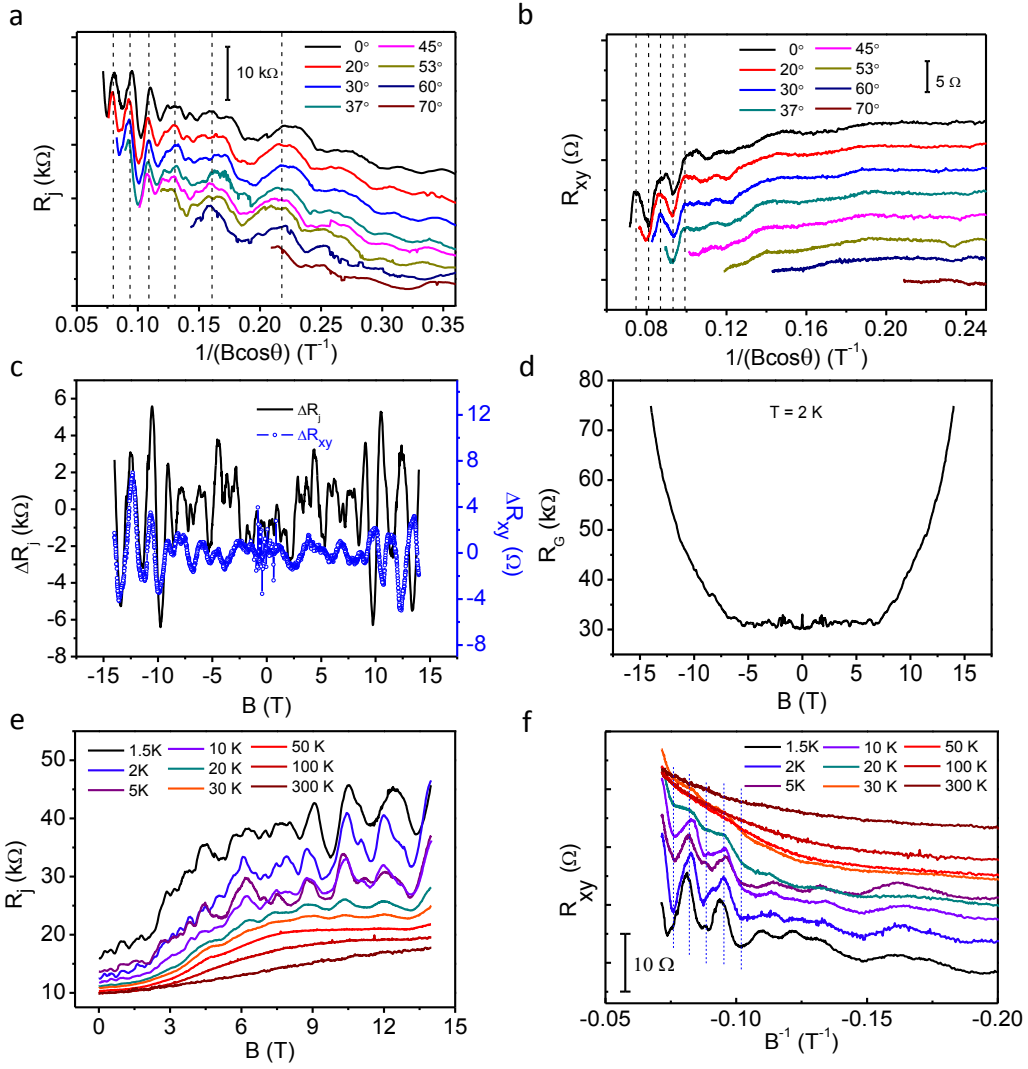

**Figure 4 | Magnetotransport properties of the device.** (a) $R_j$ and (b) $R_{xy}$ *versus* $1/B_\perp = 1/(B\cos\theta)$ for different values of $\theta$. Curves are shifted for clarity. (c) $\Delta R_j$ and $\Delta R_{xy}$ as a function of magnetic field $B$ by subtracting the background. (a), (b) and (c) were obtained at 1.5 K. (d) The resistance of graphene $R_G$ *versus* the magnetic field $B$ measured with the two-probe method at 2 K with $V_g = 0$ V. (e) $R_j$ as a function of $B$ at temperatures from 1.5 K to 300 K. (f) $R_{xy}$ as a function of $1/B$ at temperatures from 1.5 K to 300 K. The dashed lines indicate that the oscillation peaks and valleys at different temperatures line up.

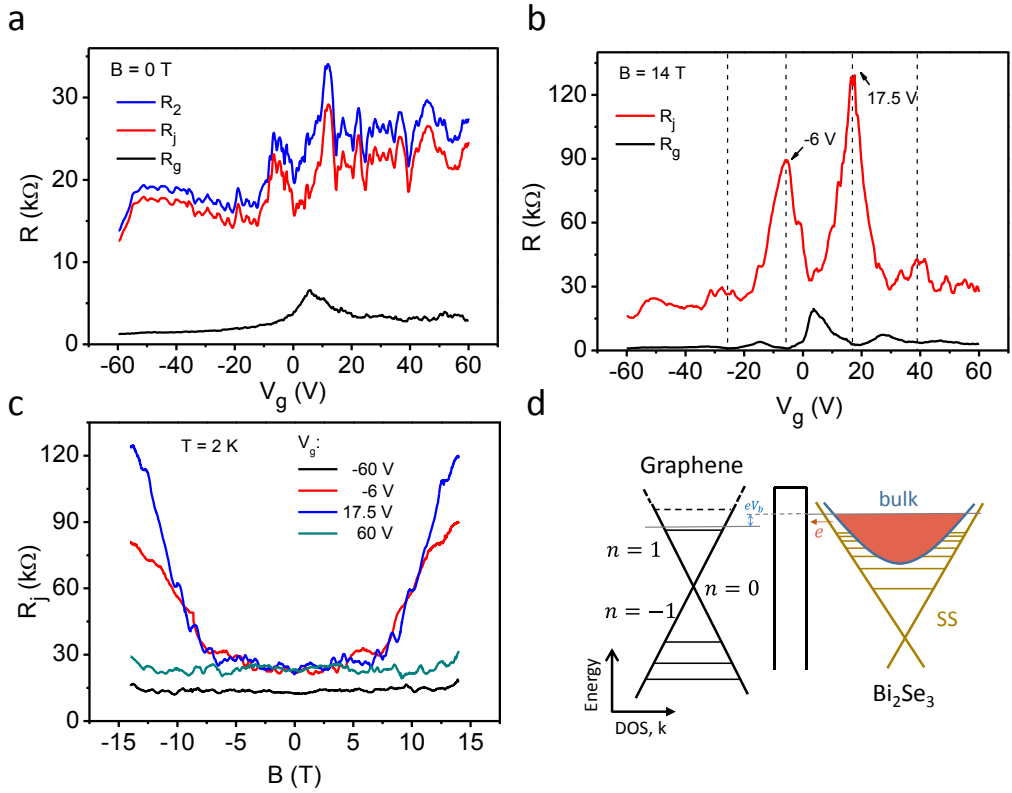

**Figure 5 | Gate-modulated tunneling transport at 2 K.** (a) Resistance *versus* gate voltage without magnetic field. (b) Transfer curves of $R_j$ and $R_g$ under $B = 14$ T. The black dashed lines are guided for eyes to show the alignment. (c) The junction resistance as a function of magnetic field under $V_g$ = -60 V, -6 V, 17.5 V, and 60 V, respectively. (d) Schematic band diagram of the tunneling process under magnetic field. Quantized LLs are formed in both graphene and $Bi_2Se_3$ under high magnetic field. The Fermi level of $Bi_2Se_3$ is almost pinned, while the Fermi energy of graphene can be tuned effectively by applying back gate voltage. Both surface and 3D bulk electrons of $Bi_2Se_3$ tunnel into graphene with a bias voltage $V_b$. When the Fermi level in graphene is located between the LLs, no empty state is available for tunneling. As the Fermi level enters the next new LL, the tunneling current recovers.

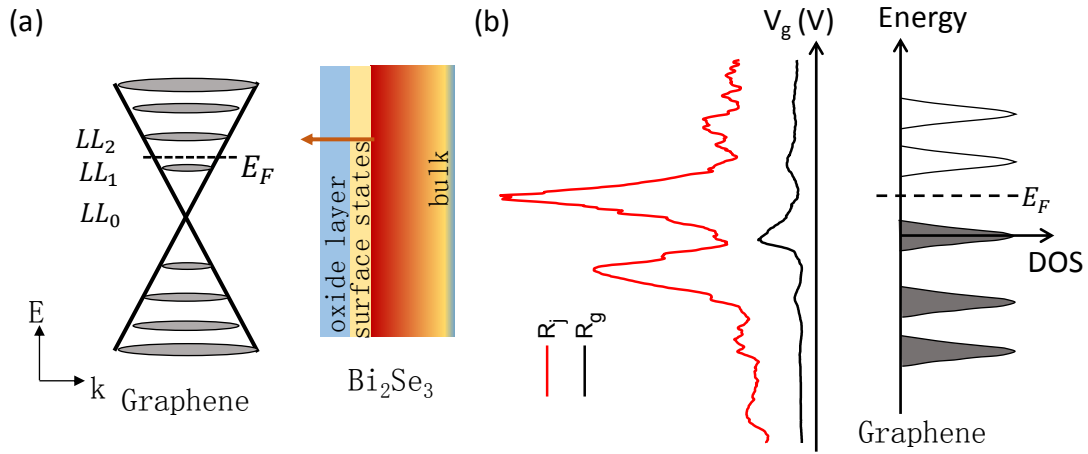

**Figure 6 | Schematic diagrams of the tunneling process in comparison with the experimental data.** (a) As the Fermi level of $Bi_2Se_3$ is almost pinned, the gate voltage will tune the Fermi level of graphene across its LLs. Electrons from both surface and bulk states of $Bi_2Se_3$ can tunnel into graphene. (b) The experimental data extracted from Figure 5b is presented to compare with the DOS structure of graphene under magnetic field of 14 T.

# Supporting Information

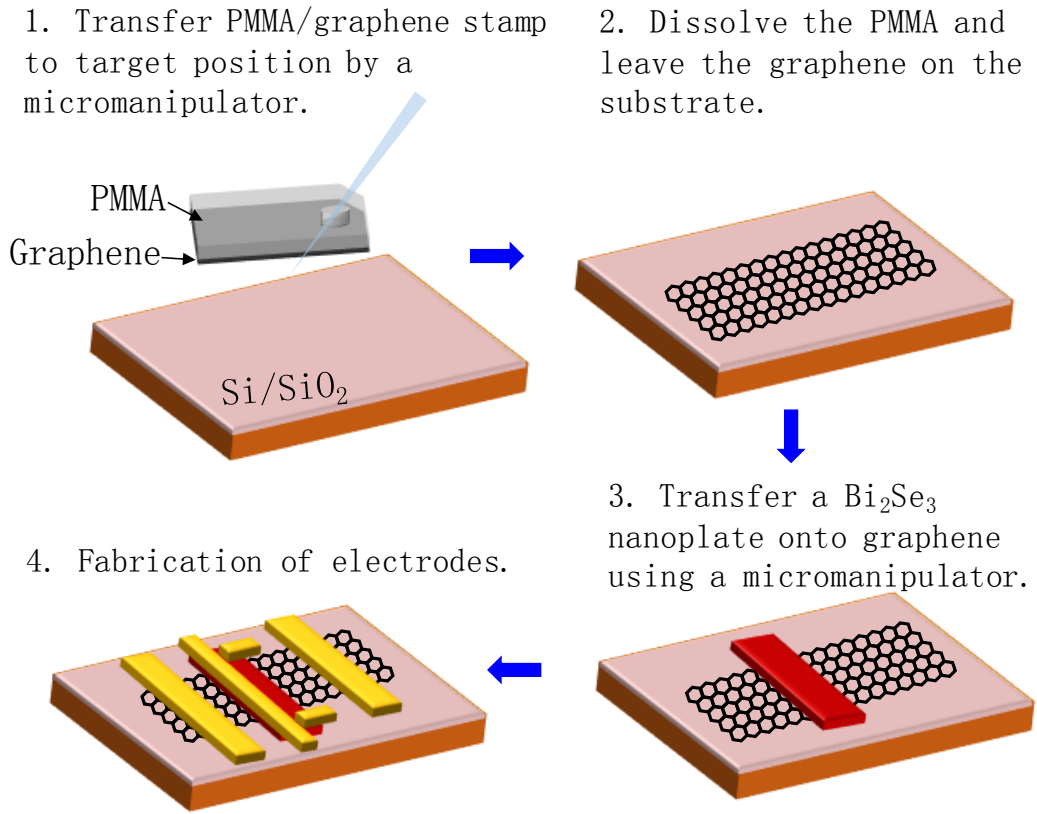

**Figure S1 | Fabrication processes of the devices.** The sketches illustrate the fabrication processes of the vertical graphene/Bi2Se3 heterostructures. See the Methods section in the main text for the detailed description.

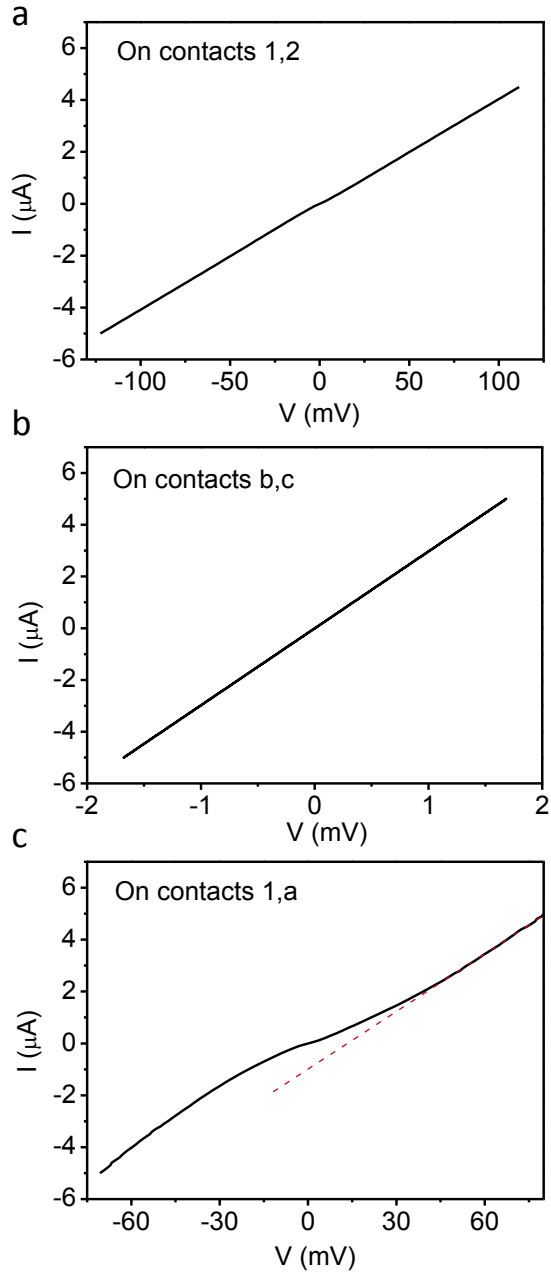

**Figure S2 | Two-terminal current-voltage curves at 2 K and 14 T.** Linear I-V curves for (a) graphene sheet measured by electrodes 1, 2, and (b) $Bi_2Se_3$ between the electrodes b and c, which indicate the Cr/Au electrodes form ohmic contacts on both graphene and $Bi_2Se_3$. (c) Non-linear and asymmetric I-V plot for the graphene-$Bi_2Se_3$ junction arises from the potential barrier at the graphene-$Bi_2Se_3$ interface.

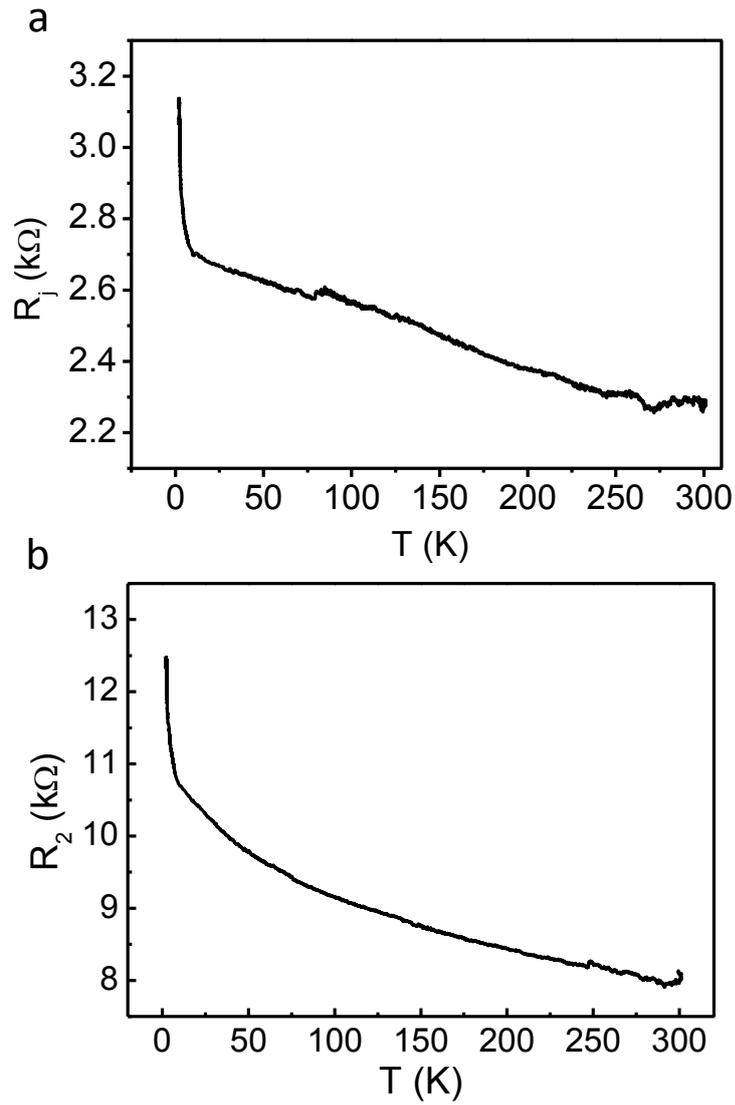

**Figure S3 | R-T behaviors of graphene-$Bi_2Se_3$ junctions.** The R-T curves in (a) measured by cross-bar configuration and (b) measured by two-probe method from two individual devices, respectively. These curves show the same tendency as those shown in Fig. 3a. It is proposed that the insulating behavior comes from the potential barrier at the interface instead of the graphene or the n-doped $Bi_2Se_3$.

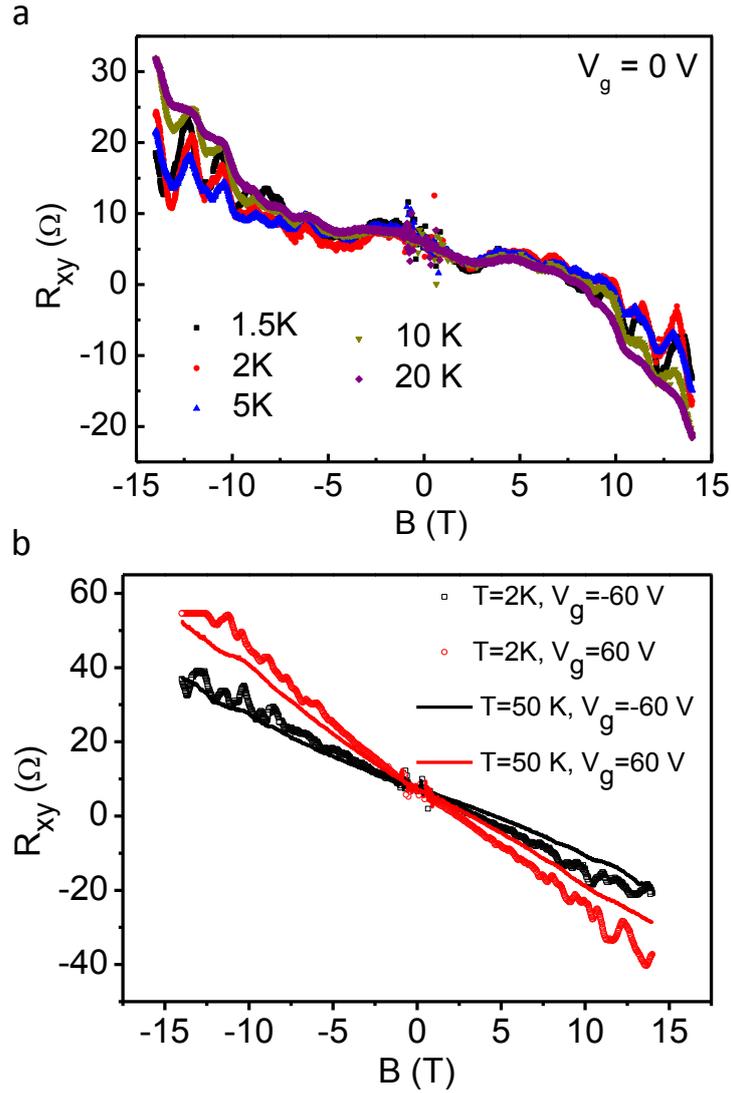

**Figure S4 | $R_{xy}$ as a function of B at different temperatures and back gate voltages.** The Hall-like resistance and longitudinal resistance are coupled due to the special electrode configuration. The negative slope of background signal indicates the major carriers are electrons in the $Bi_2Se_3$. The oscillations tend to disappear with increasing temperature, indicating the quantum origin of the oscillations. The measurement conditions, temperature and gate voltage ($V_g$) are stated in the figures. The modulation of $V_g$ up to ±60 V shows limited change in the slope of $R_{xy}$. Compared to graphene, the variation of Fermi level in $Bi_2Se_3$ is much smaller as tuning the gate voltage.

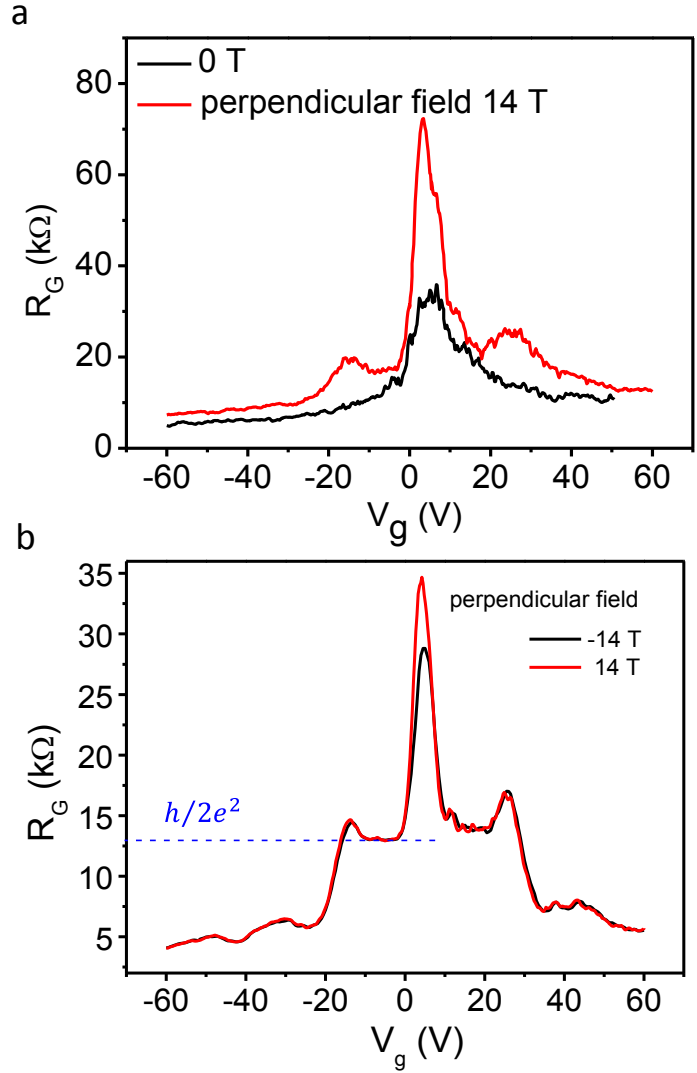

**Figure S5 | Transfer curves of graphene sheet.** Resistance of the graphene sheet versus gate voltage under perpendicular magnetic field of 14 T, -14 T and 0 T measured at 2 K using electrodes 1 and 2. (a) and (b) were measured from two different samples. Two possible reasons are responsible for the deviation of quantized values in the graphene sheet with $Bi_2Se_3$ on top. The first one is the parallel conduction of $Bi_2Se_3$. The other one is the backscattering of edge states in graphene *via* the bulk states of $Bi_2Se_3$. Nevertheless, the quantized conductance ($2\,e^2/h$) of graphene can still be observed in (b).

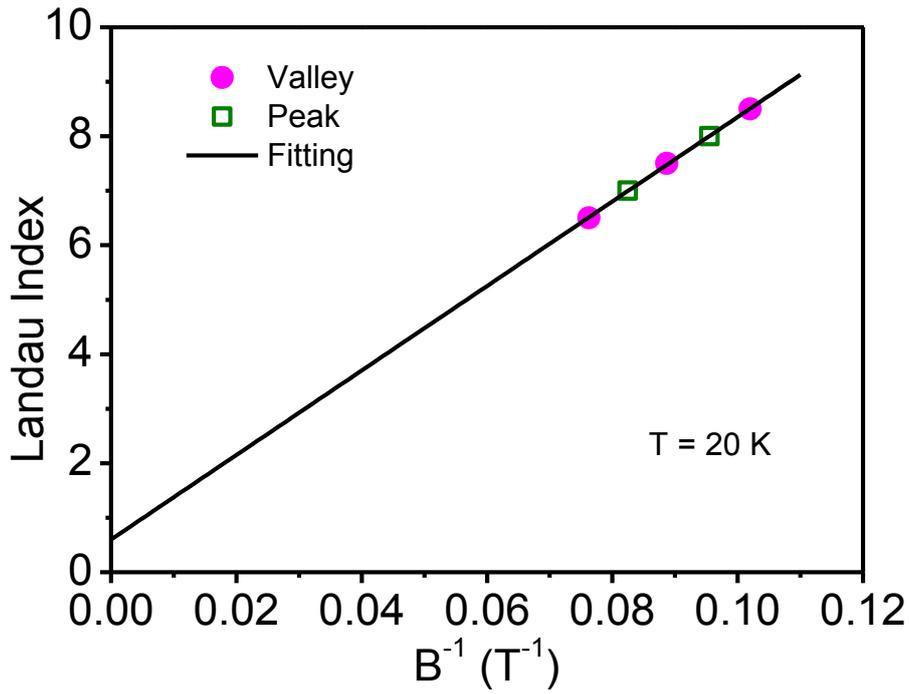

**Figure S6 | Landau fan diagram of Bi$_2$Se$_3$.** Landau fan diagram at T = 20 K extracted from R$_{xy}$ in Fig. 4f. Valley positions (closed circle) and Peak positions (open square) correspond to half-integer and integer, respectively. The rough linear fitting of the Landau index *vs.* B$^{-1}$ gives an estimation of the intercept about 0.6. A Berry phase of β = 0.8π is derived according to the Onsager relationship. The 2D carrier concentration n = $1.87 \times 10^{12}\ cm^{-2}$ is obtained from the oscillation frequency.

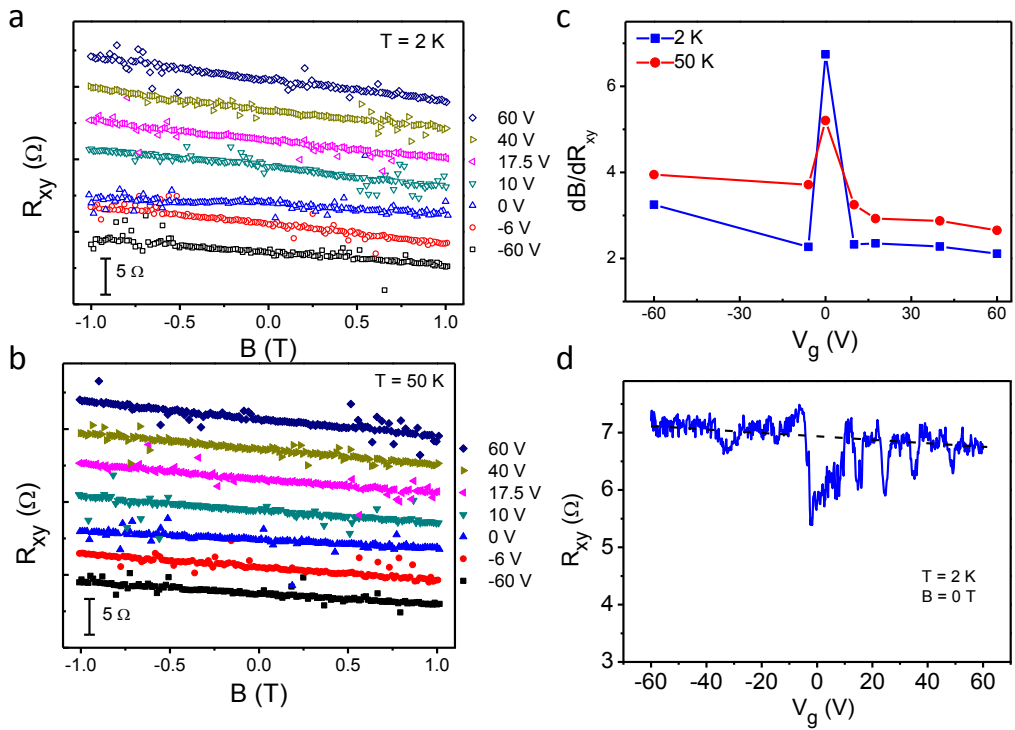

**Figure S7 | $R_{xy}$ under low magnetic field.** $R_{xy}$ versus magnetic field B from -1 to 1 T with $V_g$ = -60 V, -6 V, 0 V, 10 V, 17.5 V, 40 V, 60 V (from bottom to top) at (a) 2 K and (b) 50 K. Curves are shifted for clarity. (c) shows the $dB/dR_{xy}$ related to the carrier concentration in $Bi_2Se_3$ as a function of $V_g$ at temperatures of 2 K and 50 K. (d) $R_{xy}$ versus $V_g$ at T = 2 K and B = 0 T. The black dashed line is a visual guide.

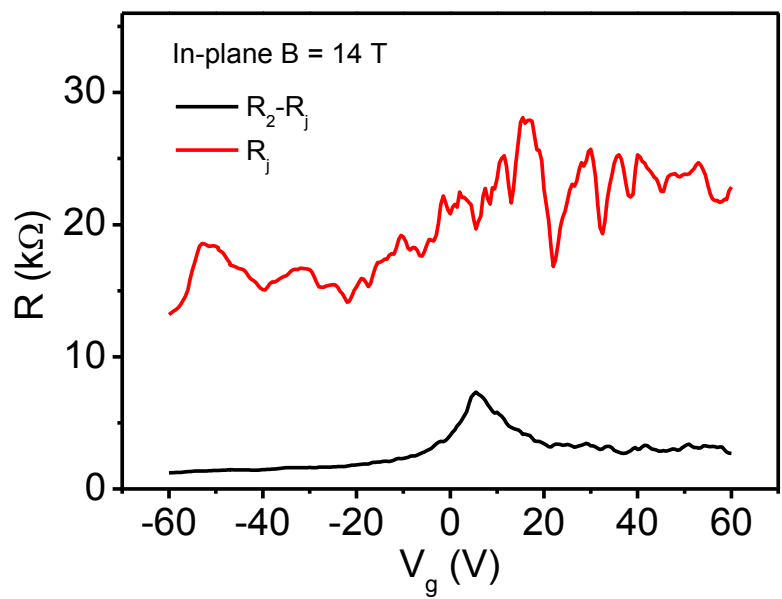

**Figure S8 | Tunneling transport under in-plane magnetic field.** Resistance $R_j$ and $(R_2-R_j)$ versus gate voltage $V_g$ plotted with an in-plane magnetic field of 14 T at 2 K. The shapes of the curves are similar with that measured under 0T as presented in Fig. 5a, indicating the 2D nature of transport.

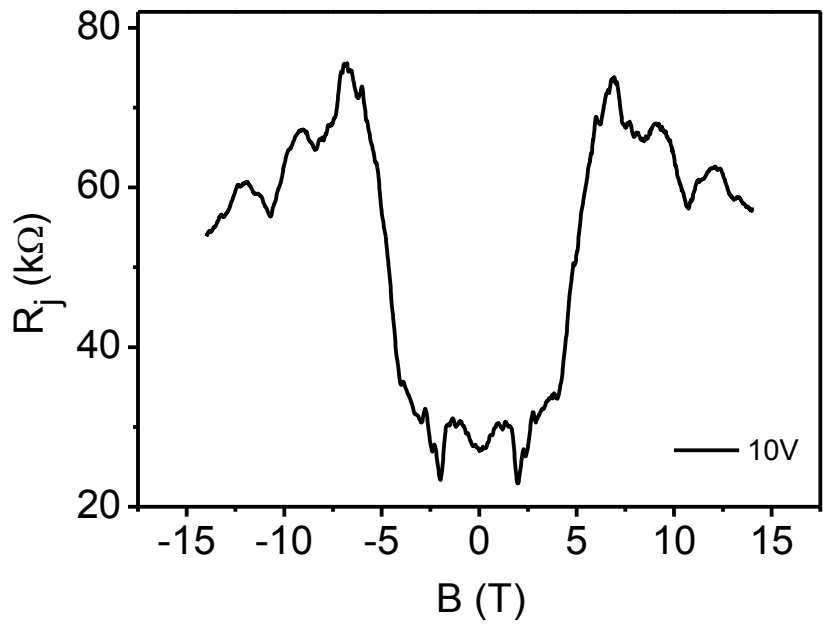

**Figure S9 | The junction resistance under perpendicular magnetic field.** $R_j$ as a function of B with $V_g = 10$ V at 2 K. As the available DOS can also be increased by enhancing the degeneracy of the LLs with increasing magnetic field, the resistance shows attenuation behavior under high magnetic field (*i.e.*, $|B| > 7$ T).

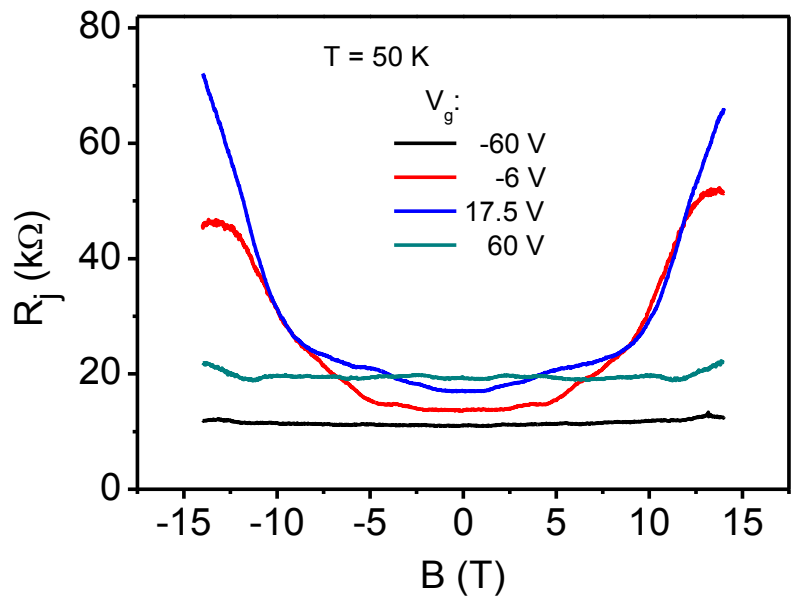

**Figure S10 | The junction resistance as a function of magnetic field at different gate voltages.** $R_j$ versus B at 50 K with $V_g$ = -60 V, -6 V, 17.5 V and 60 V.